\documentclass[twocolumn,amsmath,floatfix,nofootinbib, amssymb,pre]{revtex4}

\usepackage[]{graphicx} 
\usepackage{amsmath}
\begin{document}

\title{Electrostatic contribution to colloidal solvation in terms of the self-energy modified Boltzmann distribution}

\author{Hiroshi Frusawa}
\email{frusawa.hiroshi@kochi-tech.ac.jp}

\affiliation{Laboratory of Statistical Physics, Kochi University of Technology, Tosa-Yamada, Kochi 782-8502, Japan.}

\date{\today}

\begin{abstract}
Electrostatic interactions make a large contribution to solvation free energy in ionic fluids such as electrolytes and colloidal dispersions. The electrostatic contribution to solvation free energy has been ascribed to the self-energy of a charged particle. Here we apply a variational field theory based on lower bound inequality to the inhomogeneous fluids of one-component charged hard-spheres, thereby verifying that the self-energy is given by the difference between the total correlation function and direct correlation function. Based on the knowledge of the liquid state theory, the self-energy specified in this study not only relates a direct correlation function to the Gaussian smearing of each charged sphere, but also provides the electrostatic contribution to solvation free energy that shows good agreement with simulation results. Furthermore, the Ornstein-Zernike equation leads to a new set of generalized Debye-H\"uckel equations reflecting the Gaussian distributed charges.
\end{abstract}


\maketitle
 \pagestyle{empty}
\section{Introduction}

Electrostatic correlations in aqueous solutions including charged interfaces are important in a wide range of applications such as biological macromolecules, colloidal suspensions and ionic liquids, particularly because of the long-range nature that significantly affects many chemical and physical properties of interfaces [1-9].
The study of colloidal dispersions and electrolytes has thus received significant interest because of its central relevance to a variety of technological applications, such as in biotechnology and the food industry.
Among the various issues related to the fundamental properties of ionic solutions, we focus on the thermodynamics of ion solvation that underlies a variety of important processes, ranging from biological structures and processes to self-assemblies in soft matter systems [8-18]. 
For instance, heterogeneous assemblies are formed due to solvation of small ions and charged macromolecules such as DNA and proteins [2]. 

A vast body of theoretical literature exists for ion solvation in single-component liquids [8, 9].
Recently, the composition dependence of ion solvation in liquid mixtures was also addressed because it has been found that the solvation free energy of salt ions can significantly affect the phase behavior and interfacial properties of liquid mixtures [10-12].
Variational Gaussian approximation within a field-theoretic framework [13-17, 19] is one of the promising theories to determine electrostatic contribution to solvation free energy [13-17]. 
This method provides the Boltzmann distribution of equilibrium density whose weight is modified by the self-energy [13-17, 19].
The self-energy of an ion includes both the Born solvation effects due to a spatially varying dielectric medium and electrostatic effects [13-17];
the latter yields the ionic solvation free energy that depends on the ion concentration, valency, and the dielectric permittivity [13-18].

The variational Gaussian theory [13-17, 19] provides a self-consistent set of two equations that determines the self-energy.
One is the the self-energy modified Poisson-Boltzmann equation that goes beyond the classical Poisson-Boltzmann theory due to the inclusions of electrostatic correlations in the self-energy to improve the approximation of the mean-force potential [13-17].
The other is a generalized Debye-H\"uckel (DH) equation where the screening length spatially varies [2, 13-17, 19-26].
The generalized DH equation for obtaining the self-energy can be traced back to the inhomogeneous Ornstein-Zernike (OZ) equation [2, 20, 21, 23], according to an alternative derivation based on density functional formulation of the random phase approximation (RPA) [20, 21].
This finding opes up the possibility of developing a hybrid framework of the above self-consistent theory and the liquid state theory [1, 6, 7].

In this paper, we have advanced the above self-consistent field theory for ionic solvation.
Our formulation is based on the extension of variational field theory using the lower bound inequality [1, 27-29] to inhomogeneous fluids.
We have obtained the self-energy and its associated self-consistent equations that are more intimately connected with the liquid state theory than those of the previous formulations [13-17, 20, 21].
Accordingly, our hybrid framework based on the variational Gaussian theory (or the RPA for inhomogeneous fluids) creates a bridge between the liquid state theory and the previous findings [18, 27, 29-38] that the charge smearing (or the cut-off interaction potential) can virtually extend the scope of tractable Coulomb systems, going beyond the conventional limit of the RPA [1].
We thus obtain a simple form of the self-energy of a highly charged sphere, borrowing from the knowledge of the liquid state theory  [1, 6, 7, 30-38].
This expression does not only relate a physical picture of colloidal solvation to that known in strongly-coupled plasmas [4, 5, 30-38], but also allows us to evaluate electrostatic contribution to colloidal solvation free energy, which is in good agreement with simulation results [18] as shown below.

The remainder of this paper is organized as follows.
In section II, we define key quantities, the effective diameter $d$ and the self-energy $u({\bf r})$ per particle in the one-component charged hard-sphere system.
Section III summarizes our main results through providing a set of self-consistent equations, an equivalent to the inhomogeneous OZ equation, that determines the self-energy. 
In section IV, the validity of the self-energy formulated in this paper is assessed by seeing correspondences with simulation results of electrostatic contribution to colloidal solvation free energy, as well as with the previous formulation.
In section V, we describe the detailed derivation of $u({\bf r})$ in terms of the self-energy modified Boltzmann distribution and the Gibbs-Bogoliubov variational method.
Section VI reviews the structural outline of our hybrid theory based on both the variational Gaussian theory and the liquid state theory, and of the relation with simulation results.
Finally, concluding remarks are made in section VII.

\section{The one-component charged hard sphere (OCCH) system}
We consider the one-component charged hard-sphere (OCCH) system with its density slowly varying due to the presence of an external field.
In the highly charged systems of the OCCH, it is necessary to introduce the effective diameter $d$ of charged spheres with its actual diameter $\sigma$ and valence $q\gg 1$.
We also apply the Gaussian smearing of each charged sphere to the highly charged systems using $d$, which corresponds to a method of regularization that makes an agreement with simulation results as seen below.

\subsection{Effective diameter $d$}
In the OCCH system, the bare interaction potential $v({\bf r})$ between charged spheres consists of two parts:
$v({\bf r})=v_h({\bf r})+v_e({\bf r})$ with $v_h({\bf r})$ and $v_e({\bf r})$ denoting the hard-core and electrostatic interaction potentials.
The hard-core interaction potential $v_h({\bf r})$ is expressed as $v_h({\bf r})=\infty\,(r\leq\sigma)$ and $v_h({\bf r})=0\,(r>\sigma)$ using the sphere diameter $\sigma$ as well as the distance $r\equiv |{\bf r}|$ between charged spheres.
On the other hand, the bare electrostatic potential $v_e({\bf r})$, representing the strength of interactions between charged spheres of valence $q$ in a dielectric medium of permittivity $\epsilon$, is written as $v_e({\bf r})=q^2l_B/r$ where $l_B=e^2/(\epsilon k_BT)$ denotes the Bjerrum length, the length at which the bare electrostatic interaction between two monovalent ions is exactly $k_BT$.

The effective diameter $d$ is defined by
\begin{equation}
g({\bf r})=0\qquad(r\leq d),
\label{d definition}
\end{equation}
using the pair correlation function $g({\bf r})$ with distance $r\equiv |{\bf r}|$, and varies according to the Coulomb systems considered.
In the case of highly charged spheres, we can identify $d$ with the Wigner-Seitz radius: $d$ is determined using an electrical neutrality condition such that
\begin{equation}
\frac{4\pi}{3}\overline{\rho}d^3=q,
\label{neutrality}
\end{equation}
with $\overline{\rho}$ being the background charge density.
Note that the condition $d>\sigma$ is, strictly speaking, to be imposed on the effective diameter $d$ so that $d$ may be equated with the Wigner-Seitz radius [36];
otherwise, it is appropriate to take $d$ as a constant (i.e., $d=\sigma$) because hard-core repulsion given by $v_h({\bf r})$ leads to $g({\bf r})=0\,(r\leq \sigma)$ in the OCCH system.

\subsection{Self-energy $u({\bf r})$}
Let $G_0({\bf r}-{\bf r}')$ and $G({\bf r}-{\bf r}')$ be free and dressed propagators that satisfy the Poisson equation and a generalized set of the DH equations, respectively, for weakly-charged small ions (i.e., $q\sim1$ and $\sigma\rightarrow 0$). 
A variety of theoretical studies on ionic solvation has demonstrated that the self-energy defined by the difference between two propagators is equal to the ionic solvation free energy [13-17].
Without considering spatial dependence of dielectric permittivity (i.e. $\epsilon=\mathrm{const}.$), electrostatic contribution to the self-energy $u({\bf r})$ of a charged sphere depends on the position ${\bf r}$ due to the inhomogeneity of the present system and is simply given by [13-17]
\begin{equation}
u({\bf r})=\frac{1}{2}\lim_{{\bf r}\rightarrow{\bf r}'}\left\{
G({\bf r}-{\bf r}')-G_0({\bf r}-{\bf r}')
\right\},
\label{self2}
\end{equation}
which is similar to the conventional DH theory.
It is noted here that all energy values, including the interaction potential and free energy as well as the above self-energy, are given in the $k_BT$-unit.

We further introduce the Gaussian distribution function, $f_d({\bf r})=e^{-(\alpha r)^2/d^2}/\{\pi^{3/2}(d/\alpha)^3\}$, that will appears in the resulting equations with respect to $G_0({\bf r}-{\bf r}')$ and $G({\bf r}-{\bf r}')$. 
As will be described after eqs. (\ref{erf potential}) and (\ref{mean density}), it has been demonstrated [13, 18, 27, 30-33, 36] that the Gaussian smearing of each charged sphere due to $f_d({\bf r})$ allows us to investigate strongly-coupled Coulomb systems.
In what follows, we adopt $\alpha=1.08$, one of the values proposed by the previous theories [13, 18, 27, 30-33, 36], because the present choice of $\alpha=1.08$ has been found to provide the precise internal energy in the strong coupling regime of the uniform one-component plasma [30], as will be detailed after eq. (\ref{erf potential}).
In addition, the Gaussian smearing includes the description of weakly charged point particles (i.e., $q\sim1$ and $\sigma\rightarrow 0$) because the distribution function $f_d({\bf r})$ is reduced to the Dirac delta function: $\lim_{d\rightarrow 0}f_d({\bf r}-{\bf r}')=\delta({\bf r}-{\bf r}')$, thereby recovering the DH equation previously generalized [11-26] as confirmed below.

\section{Our results}
\subsection{The Debye-H\"uckel (DH) equations generalized using the Gaussian distribution function $f_d({\bf r})$}
The generalized DH equation set, which will be derived below, consists of three equations.
The free propagator $G_0({\bf r}-{\bf r}')$ satisfies
\begin{equation}
\nabla^2G_0({\bf r}-{\bf r}')=
-4\pi l_Bq^2f_d({\bf r}-{\bf r}').
\label{bare}
\end{equation}
Equation (\ref{bare}) is reduced to the Poisson equation of point charges due to $f_d({\bf r}-{\bf r}')\rightarrow\delta({\bf r}-{\bf r}')$.

The dressed propagator $G({\bf r}-{\bf r}')$, on the other hand, obeys
\begin{eqnarray}
\nabla^2G({\bf r}-{\bf r}')=
-4\pi l_Bq^2\left[
f_d({\bf r}-{\bf r}')-\rho^*({\bf r})
\right]
\label{GDH1}
\end{eqnarray}
for $|{\bf r}-{\bf r}'|\leq d$ and
\begin{eqnarray}
\nabla^2G({\bf r}-{\bf r}')=
\int d{\bf r}"f_d({\bf r}-{\bf r}'')
\kappa^2({\bf r}'')G({\bf r}''-{\bf r}')
\label{GDH2}
\end{eqnarray}
for $|{\bf r}-{\bf r}'|> d$.
In eq. (\ref{GDH2}), a spatially dependent screening length, $\kappa^{-1}({\bf r})=\{4\pi l_Bq^2\rho^*({\bf r})\}^{-1/2}$, has been introduced using an inhomogeneous density distribution $\rho^*({\bf r})$ specified below. 
The second term ($4\pi l_Bq^2\rho^*({\bf r})$) on the right hand side (rhs) of eq. (\ref{GDH1}) corresponds to the hole term [20, 22, 24, 25] representing the exclusion area of $|{\bf r}-{\bf r}'|\leq d$.
In the limit of $\lim_{d\rightarrow 0}f_d({\bf r}-{\bf r}')=\delta({\bf r}-{\bf r}')$, we find that eq. (\ref{GDH2}) is reduced to a generalized DH equation previously used [11-26].

Combination of the self-energy, expressed by eq. (\ref{self2}) and a generalized set of eqs. (\ref{bare}) to (\ref{GDH2}), forms the basis of our results.
The results obtained in this study will be compared with previous RPA theories [13-18, 20, 21] in more detail, after clarifying the connection with the OZ equation.

\subsection{Connection with the liquid state theory}
We focus on the typical correlation functions of concern in the liquid state theory [1]: the direct correlation function (DCF), $c({\bf r}-{\bf r}';\rho^*)$, and the total correlation function, $h({\bf r}-{\bf r}';\rho^*)=g({\bf r}-{\bf r}';\rho^*)-1$, as functions of an inhomogeneous density distribution $\rho^*$.
These correlation functions are related to each other through the OZ equation.
We will verify below that the above set of eqs. (\ref{bare}) to (\ref{GDH2}) were obtained from combining the OZ equation and the results of $G_0({\bf r}-{\bf r}')=-c({\bf r}-{\bf r}';\rho^*)$ and $G({\bf r}-{\bf r}')=-h({\bf r}-{\bf r}';\rho^*)$.

Before proceeding to the OZ equation, we need to select the concrete form of the DCF among various expressions so that the strongly-coupled OCCH system  [4, 5, 31-38] may be described precisely.
Hence, we have adopted the following DCF:
\begin{flalign}
-c({\bf r};\rho^*)&=\int d{\bf r}'\frac{l_Bq^2}{|{\bf r}-{\bf r}'|}f_d({\bf r}')=\frac{l_Bq^2}{d}\widetilde{v}_L(x),
\label{erf potential}
\end{flalign}
which has been demonstrated to be available not only for the one-component plasma but also for the OCCH system in the strong coupling regime of $q^2l_B/\sigma\gg 1$ to satisfy $d>\sigma$ [4, 5, 30-38]; otherwise, we need to decompose the DCF into the hard-core and Coulomb contributions [1].
In eq. (\ref{erf potential}), the bare electrostatic potential ($\sim 1/r$) is modified using the Gaussian distribution function $f_d({\bf r})$, and the second equation of eq. (\ref{erf potential}) introduces the function of $\widetilde{v}_L(x)=\mathrm{erf}(\alpha x)/x\,(x\equiv r/d)$ with which the long-range part of the Coulomb interaction potential is represented as $(q^2l_B/d)\widetilde{v}_L(x)$, and we have set $\alpha=1.08$ so that we may use a well-known form of the DCF relevant at strong coupling [30].
It is to be noted that the internal energy of the one-component plasma, obtained using this DCF, or eq. (\ref{erf potential}), exhibits an error of less than $0.8\%$ in the strong coupling regime [30, 38].

With the use of the above expression (\ref{erf potential}), the self-energy $u({\bf r})$ given by eq. (\ref{self2}) reads
\begin{flalign}
u({\bf r})&=\frac{1}{2}\left\{
-h(0;\rho^*)+c(0;\rho^*)\right\}
=\frac{1}{2}\left(1-\frac{l_Bq^2}{d}\widetilde{v}_L(0)\right)\nonumber\\
&\approx0.5-0.61\frac{l_Bq^2}{d},
\label{self1}
\end{flalign}
using $h(0)=-1$ and $\widetilde{v}_L(0)\approx 1.22$.
The simple form in the second line of eq. (\ref{self1}) provides a quadratic function of $q$ with $d$ treated as a fitting parameter, which will be fitted to the simulation results of electrostatic contribution to colloidal solvation free energy; consequently, we will obtain an optimized diameter $d^*$ that can be identified with the Wigner-Seitz radius in a quantitative sense.

Let us now apply the Laplacian operator to both $c({\bf r}-{\bf r}')$ and $h({\bf r}-{\bf r}')$ using the above expression (\ref{erf potential}) as well as the OZ equation: 
\begin{flalign}
&\nabla^2c({\bf r}-{\bf r}')=-\nabla^2G_0({\bf r}-{\bf r}')=4\pi l_B f_d({\bf r}-{\bf r}')
\label{nabla c}\\
&\nabla^2h({\bf r}-{\bf r}')=-\nabla^2G({\bf r}-{\bf r}')\nonumber\\
&=\nabla^2c({\bf r}-{\bf r}')
+\int d{\bf r}"
\nabla^2c({\bf r}-{\bf r}")\rho^*({\bf r}")h({\bf r}"-{\bf r}'),
\label{trans oz}
\end{flalign}
where the last equality of eq. (\ref{trans oz}) is due to the OZ equation.

Equations (\ref{erf potential}) and (\ref{nabla c}) reveal the reason why the Gaussian smearing expressed by $f_d({\bf r}-{\bf r}')$ emerges in eq. (\ref{bare}) with respect to the free propagator $G_0$.
As for $h({\bf r}-{\bf r}')$ in eq. (\ref{trans oz}), on the other hand, we need to further impose the condition, $h({\bf r})=-1\,(r\leq d)$, on eq. (\ref{trans oz}) by definition of $d$, with which eqs. (\ref{erf potential}), (\ref{nabla c}) and (\ref{trans oz}) yield our generalized DH equation set of eqs. (\ref{GDH1}) and (\ref{GDH2}) because of $f_d({\bf r}-{\bf r}')\approx 0\,(r>d)$.

\section{Validity assessment of our results}
So far, we have provided a formal list of our results: the above concrete form (\ref{self1}) of the self energy in addition to the basic equations represented by eqs. (\ref{self2}) to (\ref{GDH2}).
The validity of our results will be assessed below, through quantitative comparison with previous simulation results [18] focusing only on electrostatic contribution to the colloidal solvation free energy;
formal comparisons between ours and previous treatments of strongly-coupled Coulomb systems [13-18, 27, 34-39] are also given below.

\subsection{Correspondence with the previous forms of the self-energy [13-18, 27, 34-39]}
We can gain another physical insight into the self-energy, other than the understanding from the general form (\ref{self2}), by rewriting the expression (\ref{self1}) of the self energy as
\begin{flalign}
u({\bf r})&=\frac{1}{2}\iint d{\bf r}'d{\bf r}''f_{\sqrt{2}d}({\bf r}-{\bf r}')f_{\sqrt{2}d}({\bf r}-{\bf r}'')\nonumber\\
&\qquad\quad\times\left\{
G(0)-\frac{l_Bq^2}{|{\bf r}'-{\bf r}''|}
\right\},
\label{self3}
\end{flalign}
where another Gaussian distribution function $f_{\sqrt{2}d}({\bf r})=e^{-(\alpha r)^2/2d^2}/\{\pi^{3/2}(\sqrt{2}d/\alpha)^3\}$ has been introduced and the constancy of $G(0)=-h(0)=1$ has also been used.
Equation (\ref{self3}) implies that the self-energy is evaluated as the interaction energy difference due to effective and bare interactions between Gaussian distributed charges inside an Onsager ball (the optimally smeared charge in an object) [27, 34-39];
incidentally, the Onsager ball model (or the ionic sphere model) with Gaussian smearing has been demonstrated to yield the internal energy in the strong coupling limit [36], which is close to the accurate lower bound, or the Lieb-Narnhofer bound  [27, 34-39].

Meanwhile, the self-energy given by eq. (\ref{self2}) has the same form as that previously used in the field-theoretic formulations [13-18], except for both the absence of the hole term on the rhs of eq. (\ref{GDH1}) and the use of the Dirac delta function instead of $f_d({\bf r})$ in eqs. (\ref{bare}) to (\ref{GDH2}). 
Equation (\ref{self3}) makes clear the difference between ours and the previous form of the self-energy:
we need to plug $\alpha=\sqrt{\pi}$ into $f_{\sqrt{2}d}({\bf r})$ and to replace $G(0)$ by $G({\bf r}'-{\bf r}'')$ in eq. (\ref{self3}) for recovering the self-energy used in this previous approach [13].

\subsection{Correspondence with simulation results [18] of colloidal solvation free energy}
\begin{figure}[hbtp]
\begin{center}
	\includegraphics[
	width=8 cm
]{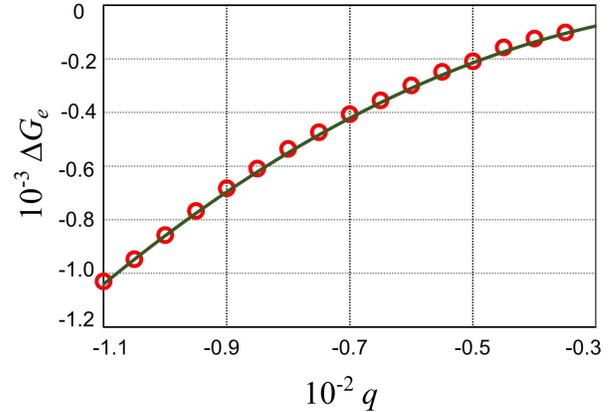}
\end{center}
\caption{Electrostatic contribution to the solvation free energy, $\Delta G_e$, as a function of the colloid charge $q$. Red circles indicate simulation results [18], and the solid line corresponds to the best fit of eq. (\ref{self1}) to these results where the effective diameter is optimized to be $d^*=5\,\mathrm{nm}$.
}
\end{figure}

Let $\Delta G_e$ be electrostatic contribution to the solvation free energy in the $k_BT$-unit due to the presence of a highly-charged colloid.
In our formulation of the OCCH system, $\Delta G_e$ is simply identified with $u({\bf r})$, the self-energy of a colloid, because the spatial variance of the dielectric constant has been ignored in this study, and the surrounding electrolyte of small ions is simply regarded as a smeared background that maintains electrical neutrality.
We compare eq. (\ref{self1}) with simulation results of $\Delta G_e$  that has been evaluated for a colloid ($\sigma = 6\,\mathrm{nm}$ and $-110\leq q<0$) surrounded by an aqueous electrolyte [18].
Note that the Bjerrum length $l_B$ in water medium at room temperature is approximately 0.7 nm, providing $l_Bq^2/\sigma>10^2$ in the range of $-110\leq q< -30$.
 Hence, this $q$-range is selected.

It has been found from the simulation results [18] that the quadratic dependence on $q$ of not only $\Delta G_e$ but also the total solvation free energy holds over a wide range of $q$, indicating that colloidal solvation is governed by electrostatic interactions and is insensitive to atomic-scale details.
Figure 1 shows the extracted results of this simulation [18] in the range of $-110\leq q< -30$.
It follows from eq. (\ref{self1}) that $u\approx 0.5-0.43q^2/d$, which provides the best fit to the simulation data using the optimized diameter of $d^*=5\,\mathrm{nm}$ in Fig. 1. 
As mentioned before, $d^*$ for highly-charged hard spheres must be equal to the Wigner-Seitz radius, or the minimum length to satisfy electrical neutrality, by definition, so that the simplified treatment of this colloidal system as an OCCH system can be justified.
Actually, the integrated charge per unit colloid vanishes around $d^*$ (see Fig. 4(A) in Ref. [18]), according to the previous result in the same simulation [18];
the thickness of the electric double layer is 2 nm and is comparable to the colloidal radius of $0.5\sigma=3\,\mathrm{nm}$ though $d^*=5\,\mathrm{nm}$ is slightly shorter than $\sigma=6\,\mathrm{nm}$.

This consistency of our approach with the simulation results suggests that electrostatic contribution to colloidal solvation free energy is ascribable to the self-energy of one colloid as a constituent of the OCCH system.
We therefore need to see the essential contribution to the self energy, going back to eq. (\ref{self3}).
Obviously, the second term on the rhs of eq. (\ref{self3}) contributes to the negative self-energy, which can be interpreted as follows:
the insertion of a single colloid causes to eliminate electrostatic interactions between smeared colloidal charges that follow a Gaussian distribution over the scale of a Wigner-Seitz cell inside which electrical neutrality is maintained, and the formation of electric double layer (or the positional rearrangement of counterions and coions) due to the existence of a highly charged colloid is represented by adjusting the Wigner-Seitz radius (i.e., $d=d^*$).
Because of the simplicity of both the physical picture and the energy form (\ref{self1}), or eq. (\ref{self3}), our approach to the evaluation of colloidal solvation free energy is expected to complement other elaborate theories, including the local molecular field theory [18, 31-33] where the existence of surrounding ions (counterions and coions) is considered explicitly and a numerical integration of its evaluation has reproduced the simulation result of $\Delta G_e$ precisely [18].

\section{Equilibrium density distribution modified by the self-energy}
\subsection{Self-energy modified Boltzmann distribution}
We have determined the self-energy given by eq. (\ref{self2}), or eq. (\ref{self1}), based on the following expression (\ref{density result2}) (or eq. (\ref{density result})) of the equilibrium density $\rho_{\mathrm{eq}}$.
Here we outline the derivation of $\rho_{\mathrm{eq}}$, in addition to $\rho^*$.

As detailed in Appendix A, our variational approach is based on the Gibbs-Bogoliubov inequality regarding the lower bound of the free energy [1, 27-29] for an inhomogeneous system whose mean-field density $\rho^*({\bf r})$ is given by
\begin{flalign}
\rho^*({\bf r})&= z\exp\left\{-\psi({\bf r})-\frac{c(0)}{2}\right\}\nonumber\\
\psi({\bf r})&=J({\bf r})-\int d{\bf r}'\rho^*({\bf r}') c({\bf r}-{\bf r}'),
\label{mean density}
\end{flalign}
with $z$ denoting the fugacity and $J({\bf r})$ an external field, which is created by a fixed charge such as a charged wall, for instance.
Equation (\ref{mean density}) corresponds to a modified Poisson-Boltzmann equation in that the bare Coulomb interaction potential $v({\bf r})$ is replaced by minus the DCF, or the long-range part of the Coulomb interaction as mentioned above (see also eq. (\ref{erf potential})) [18, 29, 31-33, 36].
It has been demonstrated that this type of the modified Poisson-Botzmann equation, or the above local molecular field theory [18, 31-33], reproduces well the simulations results in strongly-coupled Coulomb systems including the OCCH system and inhomogeneous systems of counterions dissociated from macroions [18, 31-33].

In the Gaussian approximation of the lower bound free energy regarding density fluctuations around $\rho^*$, the optimized (or maximized) lower bound is expressed using the optimized interaction potential that is identified with $-c$ (see Appendix A).
In other words, the actual grand potential $F[v]$ of the OCCH system can be approximated by $F[-c]$ with the use of the expression (\ref{erf potential}) regarding the DCF.
The equilibrium density distribution is thus obtained from the differentiation of $F[-c]$ with respect to an external field $J({\bf r})$, yielding (see below for the detailed derivation):
\begin{flalign}
&\rho_{\mathrm{eq}}({\bf r})=\rho^*({\bf r})+\frac{\rho^*({\bf r})}{2}h(0;\rho^*)\nonumber\\
&\quad\approx z\,
\exp\left[-\psi({\bf r})-\frac{1}{2}\left\{
-h(0;\rho^*)+c(0;\rho^*)\right\}\right],
\label{density result2}
\end{flalign}
where we have used the approximation, $1+h/2\approx e^{h/2}$, in the second line of the above equation.
It follows from eq. (\ref{density result2}) that the self-energy given by (\ref{self1}) has been proved.

\subsection{Detailed derivation of eq. (\ref{density result2})}
The maximum of the grand potential $F[-c]$ given by the optimized mimic interaction potential $-c({\bf r})$ consists of three parts: $F[-c]=U[\rho^*]-TS[\rho^*]+L[\rho^*]$ where $U[\rho^*]$ represents the interaction energy term in the mean-field approximation,  $-TS[\rho^*]$ is the ideal entropy term and $L[\rho^*]$ is the logarithmic correction term.
These functionals read, respectively,
\begin{flalign}
&U[\rho^*]=\frac{-1}{2}\iint d{\bf r}d{\bf r}'\rho^*({\bf r})\rho^*({\bf r}') c({\bf r}-{\bf r}';\rho^*)&
\nonumber\\
&\qquad\quad+\int d{\bf r}\,\rho^*({\bf r}) \left\{
\frac{c(0;\rho^*)}{2}+J({\bf r})-\ln z
\right\}-U_b,&
\label{energy}\\
&-TS[\rho^*]=\int d{\bf r}\left\{\rho^*({\bf r})\ln\rho^*({\bf r})-\rho^*({\bf r}) \right\},&
\label{entropy}\\
&L[\rho^*]=\frac{1}{2}\ln\,\mathrm{det}\left\{
\delta({\bf r}-{\bf r}')-\rho^*({\bf r})c({\bf r}-{\bf r}';\rho^*)
\right\},&
\label{log}
\end{flalign}
where $U_b=(\overline{\rho}^2/2)\iint d{\bf r}d{\bf r}'v_c({\bf r}-{\bf r}')$, the last term on the right hand side of eq. (\ref{energy}), arises from electrostatic interactions due to the presence of smeared background charges.
Incidentally, it has also been found based on an ionic sphere model (the Onsager ball theory [34, 36, 39]) for uniform fluids that we can obtain the same functionals as the above expressions with setting that $\alpha=1.1$, which is close to our choice of $\alpha=1.08$, from optimizing the lower bound of the internal energy with respect to the effective diameter $d$ [36], so that the internal energy may yield a similar value to the Lieb-Narnhofer bound, the lower bound in the strong coupling regime [27, 34-39].

Let $\left<\hat\rho({\bf x})\right>$ be the averaged distribution of an instantaneous density  $\hat\rho({\bf r})=\sum_{i=1}^N\delta({\bf r}-{\bf r}_i)$, that is obtained from differentiation of $F[v]$ with respect to an external field $J({\bf r})$: $\left<\hat\rho({\bf x})\right>=\delta F/\delta J$, which has been referred to as the equilibrium density $\rho_{\mathrm{eq}}({\bf r})$ [1, 6].

Considering the $\rho$-dependence of the DCF $c({\bf r}-{\bf r}';\rho^*)$, we have
\begin{flalign}
&\left.
\frac{\delta}{\delta \rho({\bf r})}
\left(
U[\rho]-TS[\rho]
\right)
\right|_{\rho=\rho^*}\nonumber\\
&=\frac{\rho^*({\bf r})}{2}\left\{
\frac{\delta c(0)}{\delta\rho}-\int_{|{\bf r}-{\bf r}'|\leq d} d{\bf r}' \rho^*({\bf r}')\frac{\delta c({\bf r}-{\bf r}')}{\delta \rho({\bf r})}
\right\},
\label{mean diff2}
\end{flalign}
where use has bee made of the approximation $\delta c/\delta\rho\approx \delta v/\delta\rho=0\,(|{\bf r}-{\bf r}'|> d)$.
Furthermore, the functional differentiation of the additional logarithmic term $L[\rho]$ with respect to $\rho$ is transformed to
\begin{flalign}
&\left.\frac{2\delta L[\rho]}{\delta\rho}\right|_{\rho=\rho^*}
&\nonumber\\
&=\int d{\bf r}\,'
\left\{
\delta({\bf r}-{\bf r}')+\rho^*({\bf r}')h({\bf r}-{\bf r}')
\right\}
&\nonumber\\
&\qquad\qquad\times\left\{-c({\bf r}-{\bf r}')-\rho^*({\bf r})\frac{\delta c({\bf r}-{\bf r}')}{\delta \rho({\bf r})}
\right\}&\nonumber\\
&=-\left\{
c(0)+\int d{\bf r}'\,\rho^*({\bf r}')h({\bf r}-{\bf r}')c({\bf r}-{\bf r}')
\right\}&\nonumber\\
&\hphantom{=}
-\left\{
\rho^*({\bf r})\frac{\delta c(0)}{\delta\rho}
+\rho^*({\bf r})\int d{\bf r}'\,\rho^*({\bf r}')h({\bf r}-{\bf r}')\frac{\delta c({\bf r}-{\bf r}')}{\delta \rho({\bf r})}
\right\}
&\nonumber\\
&\approx -\left\{
c(0)+\int d{\bf r}'\,\rho^*({\bf r}')h({\bf r}-{\bf r}')c({\bf r}-{\bf r}')
\right\}&\nonumber\\
&\hphantom{=}
-\rho^*({\bf r})\left\{
\frac{\delta c(0)}{\delta\rho}
-\int_{|{\bf r}-{\bf r}'|\leq d} d{\bf r}'\,\rho^*({\bf r}')\frac{\delta c({\bf r}-{\bf r}')}{\delta \rho({\bf r})}
\right\}
&\nonumber\\
&=-h(0)&\nonumber\\
&\hphantom{=}
-\rho^*({\bf r})\left\{
\frac{\delta c(0)}{\delta\rho}
-\int_{|{\bf r}-{\bf r}'|\leq d} d{\bf r}'\,\rho^*({\bf r}')\frac{\delta c({\bf r}-{\bf r}')}{\delta \rho({\bf r})}
\right\},
\label{ln diff}
\end{flalign}
where we have used the approximation $\delta c/\delta\rho\approx \delta v/\delta\rho=0\,(|{\bf r}-{\bf r}'|> d)$ in the third equality and the OZ equation in the last equality.

Combining eqs. (\ref{energy}) to (\ref{ln diff}),  we arrive at the result of eq. (\ref{density result2}):
\begin{align}
\rho_{\mathrm{eq}}({\bf r})&=\frac{\delta F[-c]}{\delta J}=\rho^*({\bf r})\frac{\delta J({\bf r})}{\delta J({\bf r})}
+\frac{\delta\rho^*({\bf r})}{\delta J({\bf r})}\frac{\delta L[\rho^*]}{\delta\rho^*}\nonumber\\
&=\rho^*({\bf r})+\frac{\rho^*({\bf r})}{2}h(0),
\end{align}
due to the cancellation of the terms including $\delta c/\delta\rho$.
In the second line of the above equation, we have also used the relation $\delta\rho^*/\delta J=-\rho^*$.

\subsection{Comparison with previous RPA formulations [13-21]}
While eqs. (\ref{self2}) to (\ref{GDH2}) are similar to previous ones particularly in the limit of $d\rightarrow 0$, it is evident from the expression (\ref{self1}) that our formulations have advanced the knowledge of self-energy;
actually, we have demonstrated that eq. (\ref{self1}) is available for simulation results of colloidal solvation.
However, it is still helpful for revealing the underlying physics of our results to observe, in detail, the difference between previous RPA formulations [13-21] and our results given by eqs. (\ref{self2}), (\ref{GDH1}), (\ref{GDH2}) and (\ref{self1}).

Two types of theoretical formulations are compared with our theory: density functional formulation of the RPA developed for inhomogeneous fluids [20, 21], and variational Gaussian approximation within a field-theoretical framework [13-19].
The previous formulations and proposed formulation share the same theoretical frame in the following two respects:
First, the equilibrium density distribution $\rho_{\mathrm{eq}}({\bf r})$ obeys the Boltzmann distribution that is modified by the self energy $u({\bf r})$:
\begin{flalign}
\rho_{\mathrm{eq}}({\bf r})&=z\,
e^{-\psi({\bf r})-u({\bf r})},
\label{density result}
\end{flalign}
where the potential field $\psi({\bf r})$ is determined in a self-consistent manner.
Second, the self-energy $u({\bf r})$ is evaluated using a kind of generalized DH equations since all of the theories, including the proposed formulation, take the Gaussian approximation.

A similar set of eqs. (\ref{self2}), (\ref{bare}), (\ref{GDH1}), (\ref{GDH2}), and (\ref{density result2}) has been developed through an alternative formulation to ours;
however, the interaction potential $-c({\bf r})$ described by the DCF in our equation set is replaced by the bare potential $v({\bf r})$, according to the inhomogeneous RPA [20, 21].
It follows that additional manipulations are required for treating various systems of harshly repulsive particles such as hard spheres, and that the spatial dependence in the self-energy represented by eq. (\ref{self1}) is ascribed solely to that of the total correlation function $h(0;\rho^*)$ at zero separation, due to the spatial invariance of $v(0)$ unless the dielectric permittivity spatially varies.

\section{The whole scheme of our theory}
\begin{figure*}
\begin{center}
	\includegraphics[width=13 cm
]{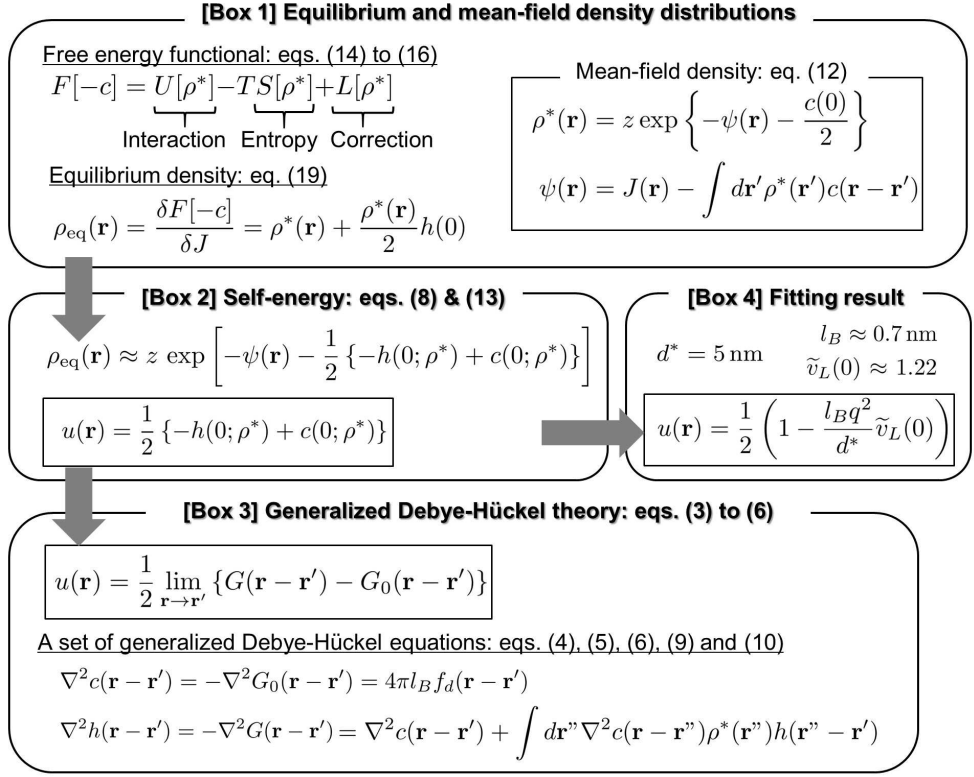}
\end{center}
\caption{The whole scheme of our theory [Box 1-3], and its comparison with simulation results [Box 4].}
\end{figure*}

In Fig. 2, there are four main boxes connected with each other through the self-energy $u({\bf r})$ and its associated equilibrium density $\rho_{\mathrm{eq}}({\bf r})$, illustrating the whole scheme of our hybrid framework based on both the variational Gaussian theory and the liquid state theory, and its comparison with simulation results:
(i) the equilibrium density $\rho_{\mathrm{eq}}$ (eq. (13)) obtained from the approximate free energy $F[-c]$ as a functional of mean-field density $\rho^*$ given by eq. (12), (ii) the self-energy modified Boltzmann distribution of $\rho_{\mathrm{eq}}$ (eq. (13)) that is expressed using the total and direct correlation functions ($h(0;\rho^*)$ and $c(0;\rho^*)$), (iii) the generalized Debye-H\"uckel theory in terms of the self-energy (eq. (3)) that is determined by solving a set of self-consistent equations (the generalized Debye-H\"uckel equation set given by eqs. (4) to (6)), and (iv) the effective diameter $d^*$ evaluated from the new form of the self-energy fitted to the simulation results of electrostatic contribution to colloidal solvation free energy.

The whole scheme of our theory is summarized in [Box 1] to [Box 3] where our formulations are reviewed in the opposite direction to that of the above sections.

{\bf [Box 1]}---
The main aim in [Box 1] is to provide the equilibrium density $\rho_{\mathrm{eq}}$, given by eq. (13), that is related to the mean-field density $\rho^*$ as well as the approximate free energy functional $F[-c]=U[\rho^*]-TS[\rho^*]+L[\rho^*]$.
The approximate density functional of $\rho^*$ is denoted by $F[-c]$ because the interaction energy $U[\rho^*]$ is described by the direct correlation function $-c({\bf r})$ as the optimized interaction potential.
The remaining contributions of $F[-c]$ consist of two parts: the ideal entropy term $-TS[\rho^*]$ and the logarithmic correction term $L[\rho^*]$ due to the Gaussian approximation of fluctuating density around $\rho^*$.
In addition, we give $\rho^*$ that is determined by a self-consistent field equation (12), which has been referred to as the modified Poisson-Boltzmann equation (see also the inner box of [Box 1]) [18, 31-33].
This equation is modified from the original Poisson-Boltzmann equation, not only in that the self-energy $-c(0)/2$ is added to the exponent of the Boltzmann distribution, but also in that the original interaction potential ($v_c({\bf r})\sim1/r$) is replaced by minus the DCF $-c({\bf r})$, the long-range part of $v_c({\bf r})$.

{\bf [Box 2]}---
The hub box in Fig. 2 is [Box 2] where we can see how the main result in this study is derived in connection with [Box 1], as well as what it yields in connection with [Box 3] and [Box 4]. 
The expression (19) written in [Box 1] is approximated by the Boltzmann distribution of $\rho_{\mathrm{eq}}$ as given by eq. (13).
Accordingly, we can verify the expression (8) of the self-energy, which is written in the inner box of [Box 2].

The self-energy in [Box 2], on the other hand, is rewritten in more tractable forms that appear in [Box 3] and [Box 4].
While eq. (3) given in [Box 3] clarifies the theoretical relationship between our result and the generalized Debye-H\"uckel formulation, a simple form of the self-energy given in [Box 4] is used for fitting of simulation results.

{\bf [Box 3]}---
In [Box 3], we can see that the self-energy expressed as eq. (8) is reduced to the conventional form (3) when regarding the difference between the total and direct correlation functions as that between dressed and free propagators ($G$ and $G_0$).
Correspondingly, we can find that the present set of generalized DH equations given by eqs. (4) to (6) is ascribed to both the inhomogeneous OZ equation and the Gaussian charge smearing that yields the DCF relevant at strong coupling;
in the limit of $\lim_{d\rightarrow 0}f_d({\bf r})=\delta({\bf r})$, the generalized DH equations previously proposed [11-26] are recovered.

{\bf [Box 4]}---
Equation (8), or the expression in the inner box of [Box 4], reveals the simple form of the self-energy that is a quadratic function of $q$.
As written in [Box 4], $\widetilde{v}_L(0)$ and $l_B$ has been found: the DCF given by eq. (7) provides that $\widetilde{v}_L(0)=1.22$, and the Bjerrum length $l_B$ in water medium at room temperature is evaluated to be 0.7 nm.
Therefore, we obtain the effective diameter $d^*$ from fitting the quadratic self-energy to the simulation results [18] regarding the electrostatic solvation energy of unit colloid as a function of the valence $q$.
The result is that $d^*=5\,\mathrm{nm}$, which we identify with the Wigner-Seitz radius.
Since the actual radius of a colloid is 3 nm in the present simulation [18], the difference between the Wigner-Seitz and actual radii (i.e., $d^*-0.5\sigma$) provides the electric double layer thickness of 2 nm, which is in good agreement with simulation results [18] in terms of  the integrated charge of unit colloid.

\section{Concluding remarks}
Thus, we have obtained the self-energy modified Boltzmann distribution of equilibrium density by extending the variational field theory [27-29] to inhomogeneous one-component fluids.
The present self-energy is simply given by the difference between $h(0;\,\rho^*)$ and $c(0;\,\rho^*)$, due to which the knowledge of the liquid state theory can be directly utilized.
The utilization of the liquid state theory naturally results in the validation of the Gaussian smearing of each charged sphere, thereby providing eq. (\ref{self1}) that is consistent with the simulation results in the strong coupling regime.
It also follows that a rephrasing of the conventional OZ equation leads to a natural extension of previously generalized DH equation [13-26] that has considered the spatial dependence of the screening length as well as the exclusion area to which the other ions are impenetrable.
Despite the intimate connection with the liquid state theory, however, there is also a difference that the approximate grand potential $F[-c]$ is a functional of not the equilibrium density $\rho_{\mathrm{eq}}$ but the mean-field density $\rho^*$, though the resulting functional form appears quite similar to the conventional density functional theory [6, 7].
It is straightforward to extend our formulation to multi-component systems;
therefore, it would be useful to apply our self-consistent field theory not only to more realistic soft matter systems such as colloids immersed in various electrolytes, but also to soft-core systems including polyelectrolyte solutions where the condition, $h(0\,\rho^*)\neq -1$, is not satisfied automatically [17, 20, 21].

\appendix
\section{The Gibbs-Bogoliubov variational method: lower bound approach}
The Gibbs-Bogoliubov inequality regarding the lower bound [1, 27-29] forms the basis of our formulations.
Let $v({\bf r})$ and $w({\bf r})$ be the bare interaction potential and a mimic interaction potential, respectively. 
The actual grand potential $F[v]$ has a lower bound, $F[w]+\Delta U[w,g]$, that depends on both a mimic interaction potential $w$ and the pair distribution function $g$ of an actual system, instead of a reference system:
\begin{flalign}
&F[w]+\Delta U[w,g]\leq F[v]\nonumber\\
&\Delta U[w,g]
=\frac{1}{2}\iint d{\bf r}d{\bf r}'\rho^*({\bf r})\rho^*({\bf r}') g({\bf r}-{\bf r}')
\nonumber\\
&\qquad\qquad\qquad\qquad\qquad\times\left\{
v({\bf r}-{\bf r}')-w({\bf r}-{\bf r}')
\right\},
\label{inequality}
\end{flalign}
where the interaction energy difference $\Delta U[w,g]$ corresponds to a correction term to a variational grand potential $F[w]$, and also the present pair correlation function $g({\bf r})$ represents density-density correlations due to density fluctuations not around the uniform density, as usual, but around the mean-field density $\rho^*({\bf r})$ given by the Boltzmann distribution (\ref{mean density}).

We can find the optimized (or maximized) lower bound by using the functional differentiation:
\begin{flalign}
\left.
\frac{\delta}{\delta w}
\left(
F[w]+\Delta U[w,g]
\right)
\right|_{w=w^*}=0.
\label{optimize}
\end{flalign}
As shown below, we perform the Gaussian approximation of $F[w]$ regarding density fluctuations around $\rho^*$, and the optimized interaction potential $w^*$ determined by eq. (\ref{optimize}) is identified with $-c$ as will be given in eq. (\ref{w-c}).

It has been shown that the grand potential $F[w]$ with an arbitrary interaction potential $w({\bf r}-{\bf r}')$ is expressed by the density functional form [27-29, 40]:
\begin{equation}
e^{-F[w]}=\int D\rho\, e^{-U_w[\rho]+TS[\rho]},
\label{density integral}
\end{equation}
where $S[\rho]$ has been given by eq. (\ref{entropy}), and the functional form of $U_w$ is the same as eq. (\ref{energy}) if only the DCF is replaced by $-w$:
\begin{flalign}
&U_w[\rho]=\frac{1}{2}\iint d{\bf r}d{\bf r}'\rho({\bf r})\rho({\bf r}') w({\bf r}-{\bf r}')&
\nonumber\\
&\qquad\quad+\int d{\bf r}\,\rho({\bf r}) \left\{
\frac{-w(0)}{2}+J({\bf r})-\ln z
\right\}-U_b,&
\label{energy2}\\
&-TS[\rho]=\int d{\bf r}\left\{\rho({\bf r})\ln\rho({\bf r})-\rho({\bf r}) \right\}.&
\label{entropy2}\\
\end{flalign}
The saddle-point equation,
\begin{flalign}
\left.
\frac{\delta}{\delta \rho({\bf r})}
\left(
U_w[\rho]-TS[\rho]
\right)
\right|_{\rho=\rho^*}=0,
\label{sp}
\end{flalign}
thus provides the mean-field density given by eq. (\ref{mean density}) [27-29, 40].
Performing the Gaussian approximation of density fluctuations around $\rho^*$ in eq. (\ref{density integral}), we have
\begin{equation}
F[w]=U_w[\rho^*]-TS[\rho^*]+L[\rho^*].
\label{fw}
\end{equation}
The variational functional $F[w]+\Delta U[w,g]$ can be maximized at the optimized interaction potential of $w^*$ (the mimic interaction potential) that is determined by eq. (\ref{optimize}).
Plugging eq. (\ref{inequality}) and (\ref{fw}) into the relation (\ref{optimize}), we have
\begin{flalign}
&h({\bf r}-{\bf r}')=-w^*({\bf r}-{\bf r}')-\int d{\bf r}''\,
w^*({\bf r}-{\bf r}'')\rho^*({\bf r}'')h({\bf r}''-{\bf r}'),&
\label{oz2}
\end{flalign}
which is nothing but the OZ equation for inhomogeneous fluids (see Ref. [29] for the detailed derivation).
Hence it has been confirmed that the optimized potential $w^*$ is identified with minus the DCF as mentioned above:
\begin{equation}
w^*({\bf r}-{\bf r}')=-c({\bf r}-{\bf r}';\rho^*),
\label{w-c}
\end{equation}
where $c({\bf r}-{\bf r}';\rho^*)$ depends on $\rho^*$ through the OZ equation for inhomogeneous fluids.

It is to be noted that $\Delta U[w^*=-c,g]$ vanishes in the mean spherical approximation where $g(v+c)\equiv 0$ [1].
We have also ignored $\Delta U[-c,g]$ even in the hypernetted chain approximation; this corresponds to the neglect of the correlation entropy difference between the mean spherical and the hypernetted chain approximations (see Ref. [28] for the details).

\end{document}